# Second-harmonic generation in bulk diamond based on inversion symmetry breaking by color centers

*Aizitiaili Abulikemu[*,†], Yuta Kainuma[‡], Toshu An[‡], and Muneaki Hase[*,†]*

[†]Department of Applied Physics, University of Tsukuba, 1-1-1 Tennodai, Tsukuba 305-8573, Japan.

[‡]School of Materials Science, Japan Advanced Institute of Science and Technology, Nomi, Ishikawa 923-1292, Japan.

**ABSTRACT:** Breaking inversion symmetry in solids plays a central role in nonlinear optics because it can change material properties—such as producing even-order nonlinear optical (NLO) effects. Although centrosymmetric diamond has been developed as photonic platforms including waveguides and light sources, the NLO effects in bulk diamond are limited to the third-order. Thus, exploiting more powerful second-order NLO effects—such as second harmonic generation (SHG)—are still challenging. Here we explore symmetry-breaking-induced second-order NLO effects in bulk diamond using the color center—nitrogen-vacancy center. Exciting with ultrashort laser pulses, SHG and third-harmonic generation (THG) are simultaneously observed at the same time, exhibiting characteristic intensity patterns depending on both the excitation fluence and the angle of light polarization. We uncovered that SHG serves as the source for THG by the cascading process. Our findings will offer SHG-based quantum sensing by diamond color centers—such as imaging of electromagnetic field by electro-optic effects on the nanofemto scale.

Natural single-crystal diamond can be produced under extremely high-pressure and high-temperature (HPHT) environment in the earth's mantle.[1] In contrast, artificial diamond crystals can be also made under the HPHT conditions[2] or using chemical vapor deposition method (CVD),[3] providing emerging power electronic and photonic applications.[4-6] Furthermore, diamond color centers, such as nitrogen-vacancy (NV) or silicon-vacancy (SiV) center are attracting considerable attention owing to their potential applications in quantum sensing,[7,8] biomedicine,[9] and quantum information sciences.[10-12] Since diamond single-crystal lattice, i.e., cubic structure, holds inversion symmetry, second-order NLO susceptibility $\chi^{(2)}$ in bulk region is strictly zero.[13] Therefore, third-order NLO effects are believed to be the lowest order over the decades. By this fundamental reason, most of the investigations on NLO effects in the bulk diamond crystals has been developed based on the third-order NLO susceptibility $\chi^{(3)}$, that governs optical Kerr effect,[14,15] two-photon absorption,[16,17] and so on. On the other side, there are a few investigations on NLO effects in nanocrystalline diamonds, where generation of SHG was based on the symmetry breaking by grain boundaries.[18,19] However, their non-uniform morphological and optical properties will hinder the potential applications of such solute nanocrystalline diamonds.[20]

The NLO processes is generally described by nonlinear polarization in nonlinear media, $\boldsymbol{P}_{NL} = \chi^{(1)}\boldsymbol{E} + \chi^{(2)}\boldsymbol{EE} + \chi^{(3)}\boldsymbol{EEE} + \cdots$, where $\chi^{(1)}$ is the linear susceptibility and $\boldsymbol{E}$ is the driving electric field.[13] To allow for the second-order perturbation available, researchers frequently use inversion symmetry breaking by the surface dangling bonds, which enables SHG and even sum-frequency generation from the surface dipole in solid-state materials, including semiconductors (Si),[21] topological insulators ($Bi_2Se_3$),[22] and transition metal dichalcogenides ($MoS_2$).[23] However, such NLO signals at a surface are very weak,[21,22] except a resonant excitation condition with narrow spectral tuning range.[23]

Here we show carefully designed NV centers in a diamond crystal can also break the inversion symmetry in the bulk region (Figure 1a, b), making it possible to realize second-order NLO processes by the bulk dipole and even quadrupole, where the second-order nonlinear polarization of $\boldsymbol{P}^{(2)}(2\omega) = \chi^{(2)}\boldsymbol{E}(\omega)\boldsymbol{E}(\omega)$ acts as the driving source, where $\omega$ is the optical carrier frequency. Illuminating by femtosecond (fs) infrared (IR) light pulses, it is revealed that harmonic signals are detected by phase-matching (wave-vector conservation) conditions in a wide range of visible wavelength without interruption of optical properties of diamond. From application point of view, this work will open new avenues to the emergence of NV diamond-based second-order nonlinear photonics. For instance, it will enable one to generate wavelength-tunable SHG from diamond colour centers, covering from the X-ray to the far-infrared regime. Moreover, other second-order nonlinear effects, such as optical rectification and electro-optic effect,[13] will also help development of diamond nonlinear photonics toward next generation of quantum sensing technologies.

All-solid-state, type-IIa, chemical vapor deposited single crystal diamond samples with (100) crystal orientation, whose impurity (nitrogen: [N] and boron: [B]) levels were [N] < 1 ppm and [B] < 0.05 ppm, respectively, was used for this study. The sample size was 3.0 mm×3.0 mm×0.3 mm (thickness). The NV diamond sample hosting about 35 nm-deep NV centers was prepared via combination of the nitrogen ions ($N^+$) implantation (30 keV at a $N^+$ dose of $5.0×10^{12}$ ions $cm^{-2}$) and high temperature annealing (1 hour at 900 °C in an argon atmosphere) techniques.[24] The corresponding estimated value of density of NV centers was ≈$2.6×10^{17}$ $cm^{-3}$, assuming a production efficiency of 10%.[24] The production of NV centers in diamond sample was confirmed by the characteristic peaks observed in the photoluminescence (PL) spectrum (Figure 1c) excited with continuous-wave (CW) laser at 532 nm. As displayed in the PL spectra, the electronic state of the NV diamond was the negatively charged state ($NV^-$) because of the observation of the characteristic broadband at ∼ 660 nm together with the much narrower zero-phonon line (ZPL) peak at ≈ 638 nm, which is stronger than that from the neutrally charged state ($NV^0$) observed at ∼ 600 nm and ≈ 575 nm, respectively.[25]

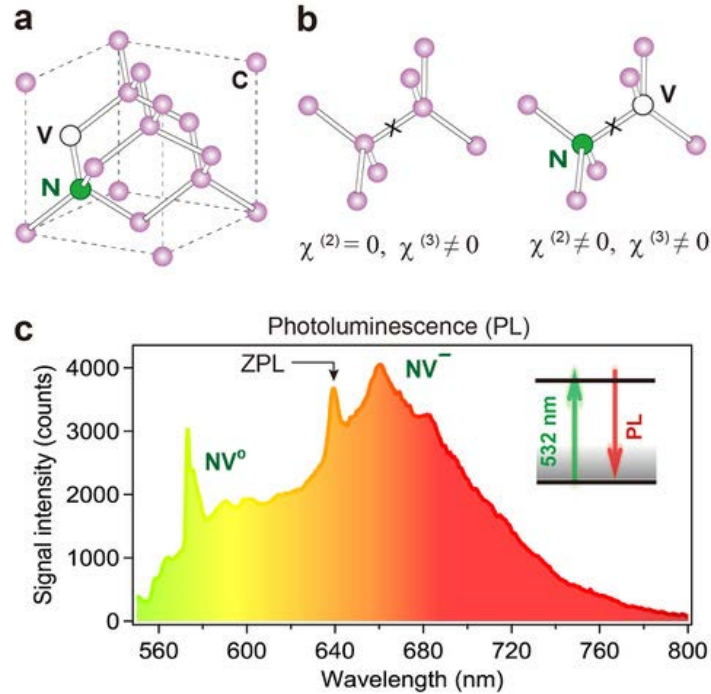

**Figure 1. Diamond crystal structure and photoluminescence spectra.** (a) The crystal structure of nitrogen (N)-vacancy (V)-introduced diamond, where carbon atoms (C) bond to four others in a cubic lattice. (b) Schematic local structure of inversion symmetry hold (left) and broken (right) diamond. The mark " × " represents the inversion center. (c) Photoluminescence (PL) spectra from the NV diamond sample excited with CW laser (532 nm) recorded at room temperature. In addition to the broad PL bands due to the NV⁻ and NV⁰ states, the narrow zero-phonon line (ZLP) appears at 638 nm overlaid on the NV⁻ band.

Nonlinear optical spectra were measured with a home-built optical spectroscopy system at normal incidence in reflection geometry and all spectra were recorded at room temperature (293 K). Ultrafast laser pulses from a mode-locked Ti:sapphire laser were amplified (800 nm; 1.55-eV and 100 kHz repetition rate) and delivered to an optical parametric amplifier to generate wavelength-tunable (from 1200 to 1600 nm) femtosecond laser pulses. The IR excitation beam spectrally cleaned through a band-pass filter, power of the exciting laser pulse is controlled by a neutral density filter, and the excitation beam was focused on the sample surface utilizing a focusing lens ($f$ = 50 mm) to a beam diameter of ≈ 20 μm. The omnidirectional emission signals were collected by a collimating objective lens to couple into an optical fibre (a core diameter of 200 μm), and finally detected by a spectrometer through a dichroic mirror. The polarization tests were performed on diamond samples using the same optical geometry and excitation fluence of $I_{ex}$ =7.6 mJ/cm², but inserting a half waveplate.

Figure 2 demonstrates observation of nonlinear emission from the NV diamond crystal compared with the direct THG from pure diamond crystal (without NV centers). For pure diamond single crystal in which there rarely exist impurities and defects, optical absorption starts from ≈ 5.5 eV (225 nm) due to the indirect band-gap along the $\Gamma - \Delta$ points or more resonant exciton (an electron-hole pair) bands at ∼ 5.4 eV (230 nm).[26] By irradiating IR laser pulses far-bellow the band-gap energy at 0.92 eV ($\lambda_{ex}$= 1350 nm), the third-order nonlinear polarization of $\boldsymbol{P^{(3)}}(3\omega) = \chi^{(3)}\boldsymbol{E}(\omega)\boldsymbol{E}(\omega)\boldsymbol{E}(\omega)$, which is controlled by $\chi^{(3)}$, plays a pivotal role to produce direct THG.[27] Based on the coupled-wave equations for THG we can estimate the value of $\chi^{(3)}$ using the measured conversion efficiency $\eta$,[18]

$$\chi^{(3)} = \frac{2\varepsilon_0 c \lambda_{ex}\sqrt{n_\omega^3 n_{3\omega}}}{3\pi L I_{ex}}\sqrt{\eta}, \qquad (1)$$

where $c$ is speed of light, $n_\omega$ and $n_{3\omega}$ are the refractive indexes for the fundamental ($\omega$) and THG ($3\omega$) frequencies, respectively, $L = \lambda_{ex}/(3n_{3\omega} - 3n_\omega)$ is the coherence length. Using Equation (1) with $\eta = 6.5\pm0.9\times10^{-4}$, we obtain $\chi^{(3)} = 1.1\times10^{-20}$ m$^2$/V$^2$ for the pure diamond crystal, which is more than one-order of magnitude larger than that obtained for pure single crystal or nanocrystalline diamonds using transmission geometry.[14,18] The larger $\chi^{(3)}$ obtained in the present study may be explained by nearly perfect phase matching in the reflection geometry.[13] Thus, the nonlinear emission spectrum of the pure diamond crystal (Figure 2a) exhibits only the direct THG signal at $\approx 450$ nm, by irradiating with IR light pulses—60 fs in duration and centered at $\lambda_{ex} \approx 1350$ nm—at a excitation fluence of $I_{ex}$= 7.6 mJ/cm$^2$ (see Figure S1a in Supporting Information). Absence of the SHG signal in the pure diamond indicates that neither the surface dipole nor bulk quadrupole contribute to the NLO process.

On the contrary, the NV color centers hosting near the surface region can break the inversion symmetry of diamond crystal lattice (Figure 1b), and it will become a potential source for the second-order NLO phenomena, such as SHG. Under the same experimental conditions, we are able to detect the SHG (at $\approx 675$ nm) in addition to the THG (at $\approx 450$ nm) from the NV diamond crystal (Figure 2b). The obtained nonlinear conversion efficiencies of SHG and THG from NV diamond are $\eta = 4.7\pm0.9\times10^{-5}$ and $\eta = 3.0\pm0.8\times10^{-5}$, respectively. When we continually varied the excitation wavelength, both the SHG and THG peak wavelengths are linearly tuned, under the constant excitation fluences of $I_{ex}$ (Figure S1b in Supporting Information). The obtained nonlinear conversion efficiency for SHG and THG shows a gradual decrease with increasing IR excitation wavelength (Figure S2 in Supporting Information). This could be correlated to a decrease of the effective pump fluence due to the larger beam spot size on the sample using the same focusing lens for the longer wavelength, and to phase mismatch due to broader bandwidth (Figure S3 in Supporting Information). Thus, the observation of SHG indicates that the introduction of NV centers could convert the surface region of bulk diamond crystal from inversion symmetric ($\chi^{(2)} = 0$) to the inversion symmetry breaking ($\chi^{(2)} \neq 0$), allowing the second-order NLO effects.

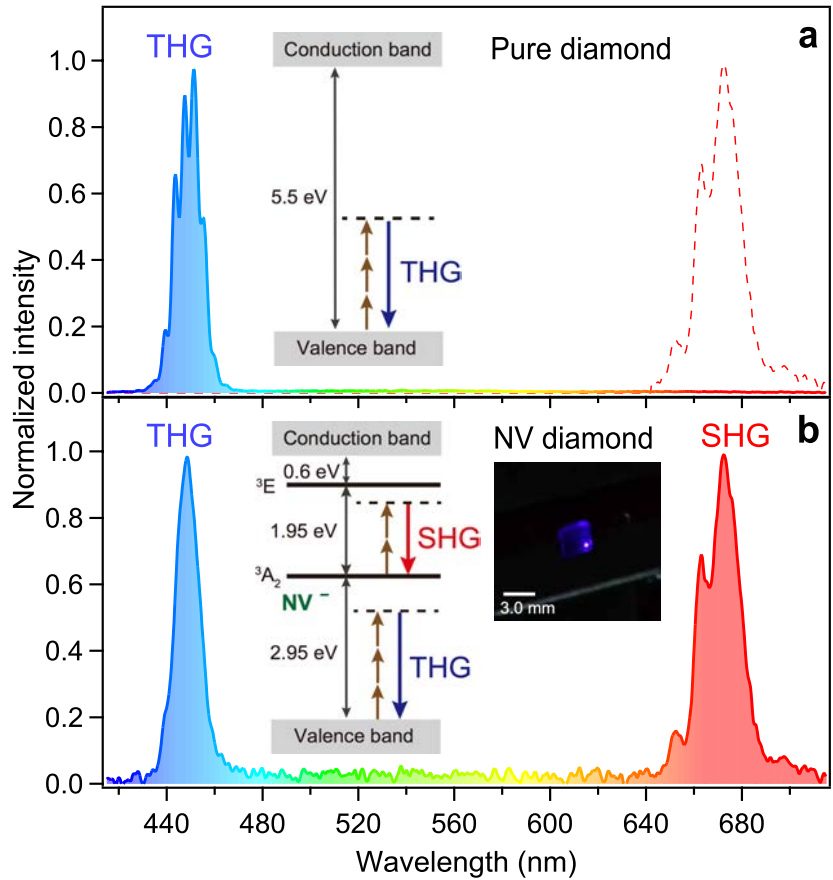

**Figure 2. Nonlinear emission spectra of diamond crystals.** (a) The nonlinear emission spectrum of the pure diamond crystal excited with an IR laser (wavelength centered at $\lambda_{ex} \approx$ 1350 nm and $I_{ex} =$ 7.6 mJ/cm$^2$). The inset represents the energy diagram for generation of the direct THG. (b) The nonlinear emission spectrum of the NV diamond crystal excited with IR laser under the same optical conditions in (a). The signal of Fig. 2a is much stronger than that in Fig. 2b, and the scaling factor between panel (a) and (b) is $\approx$ 25. The inset represents the energy diagram for generation of both SHG and THG from the NV diamond, hosting the NV- states whose ground- and excited-states are levelled by $^3A_2$ and $^3E$, respectively. The brawn arrows represent excitation photons, while the red and blue arrows represent the SHG and THG emissions, respectively. A photograph was taken during the nonlinear emission (SHG and THG) from the NV diamond crystal excited with 1350 nm pulses.

To clarify the origin of the SHG and THG from the NV diamond, Figure 3 explores the excitation fluence dependence of the signal intensities. As the excitation fluence increased from $I = 3.0$ to 13.2 mJ/cm$^2$ both the SHG and THG signals increase, but with a different manner (see Figure S4 in Supporting Information). In fact, the integrated intensities give different exponents; the SHG signal intensity is found to be proportional to $\approx I^2$, while that of the THG is to $\approx I^3$, respectively, both of which are consistent with the phenomenological relationships expected in the NLO effects.[13,28] At the lower excitation fluence than $\approx 8.3$ mJ/cm$^2$, the SHG signal intensity from NV diamond is greater than that of the THG, suggesting SHG dominates over the all-order nonlinear process when the squared electric field amplitude was less than $\approx 8.3$ mJ/cm$^2$. In contrast, at the higher fluences regions than $\approx 8.3$ mJ/cm$^2$, the THG from NV diamond grows faster than SHG. It is interesting to note that the emission intensity from the NV diamond in Figure 3 shows a similar tendency of the harmonic signals observed from superlattice-like optical materials such as a periodic LiTaO$_3$ crystal.[29] Furthermore, both the SHG and THG intensities significantly saturate at a fluence of $\approx 8.3$ mJ/cm$^2$, implying exhaust of optical transition from the limited number density of the ground states of the NV centers ($^3$A$_2$ in Figure 2b). We emphasize that the intensity variation observed in Figure 3 shows a kink at $\approx 8.3$ mJ/cm$^2$ both for the SHG and THG in NV diamond, implying a close relationship between the origins of the SHG and THG. Noteworthily, there is another possible interpretation for THG in nonlinear crystals, namely the cascaded THG ($3\omega = 2\omega + \omega$), which is governed by cascading $\chi^{(2)}$; $\chi^{(3)} = \chi^{(2)} \cdot \chi^{(2)}$, and it could give higher conversion efficiency than the direct THG.[13]

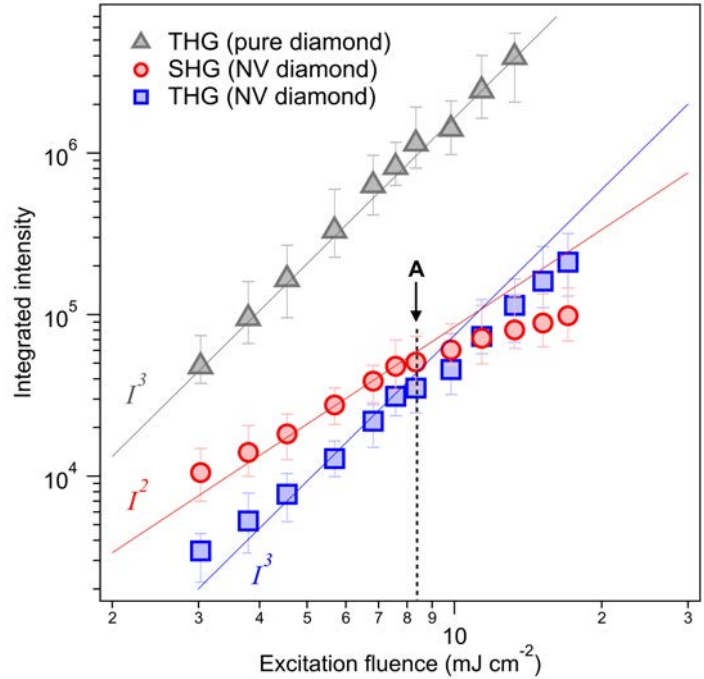

**Figure 3. Double logarithmic plot of the intensity of the harmonic signals.** Log versus Log plots of the SHG ($\approx 675$ nm) integrated intensities for NV diamond and THG ($\approx 450$ nm) for both the pure and NV diamonds as the function of excitation fluences with the fixed center wavelength at $\lambda_{ex} =$ 1350 nm; integrated from 600 nm to 750 nm for SHG and from 400 nm to 500 nm for THG. The excitation fluences dependence (exponent) is found to be ≈ 2 for SHG and ≈ 3 for THG. The arrow (A) and black dashed line represent the saturation point (a kink at 8.3 mJ/cm$^2$) of the harmonic signals for NV diamond.

To further investigate the relationship between the SHG and THG in NV diamond from the viewpoint of crystal symmetry of pure and NV diamond single crystals, we carried out the excitation polarization dependence of the intensity of harmonic signals. As illustrated for pure diamond in Figure 4a, the driving field acts directly on the atomic bonds to produce the direct THG, which is mapped by the polar plots of the direct THG intensity in Figure 4b. The normalized intensity exhibits the fourfold $O_h$ symmetry with a 45° tilt, meaning that the electronic quadrupole contribution is dominated in $\boldsymbol{P}^{(3)}(3\omega)$, which is induced by the charge distribution along the high-symmetry four equivalent bond axes of <111>, <$\bar{1}\bar{1}$1>, <$\bar{1}$1$\bar{1}$>, <1$\bar{1}\bar{1}$> in pure diamond.[30] Surprisingly, unlike for the direct THG in pure diamond, both SHG and THG from NV diamond crystal show a twofold symmetry nature, as demonstrated in Figures 4d and 4f. The polarization direction for the SHG output is nearly identical (parallel) to the *s*-polarized axis of the IR excitation light with $C_{2v}$ rotation symmetry. This result indicates that the electronic dipole contribution dominates in $\boldsymbol{P}^{(2)}(2\omega)$ over the quadrupole contribution, suggesting that the dipole moment induced by the NV center is aligned along the <001> axis (Figure 4c); corresponding to the radiation by *z*-dipole condition for the (100) face as reported for Raman scattering.[21, 30] In addition, the similarity of the polar plot of THG with that of SHG for the NV diamond is interpreted as the cascaded THG (Figure 3e), i.e., $\boldsymbol{P}^{(3)}(3\omega) = \chi^{(2)}\{\boldsymbol{E}(\omega)\boldsymbol{E}(\omega)\}\chi^{(2)}\boldsymbol{E}(\omega)$. Note that using Equation (1) we obtain $\chi^{(3)} = 2.4 \times 10^{-21}$ m$^2$/V$^2$ for the NV diamond, which is slightly smaller than that of $\chi^{(3)}$ obtained for the pure diamond crystal, and both values are similar order with previous reports.[18,31] The calculated value of $\chi^{(2)} = 4.9 \times 10^{-11}$ m/V for the NV diamond from $\chi^{(3)}$ is comparable to the $\chi^{(2)}$ values obtained for conventional nonlinear crystal such as $\beta$ -BaB$_2$O$_3$ and LiNbO$_3$.[13] Our previous work suggests that the nonlinear coefficients ($n_2$) increased as the density of NV centers increased.[15] Therefore, we would expect that the saturation power for the NV diamond crystal (8.3 mJ/cm$^2$) will be increased when we could add more NV centers, although further investigation is required by examining generation of more defects by ion irradiation without irradiation damage.

Note that the polarization direction of SHG and the cascaded THG outputs (Figures. 4d and 4f) shows a small tilt with an azimuthal angle of ≈ 15~20 degrees, possibly owing to the different phase matching conditions for SHG and the cascaded THG, in addition to more minor contributions from (i) distribution of the axes of NV centers and (ii) misalignment of the surface-orientation of the NV diamond sample, both of which would introduce some variations in the tilt of the cascaded THG polarization axis.

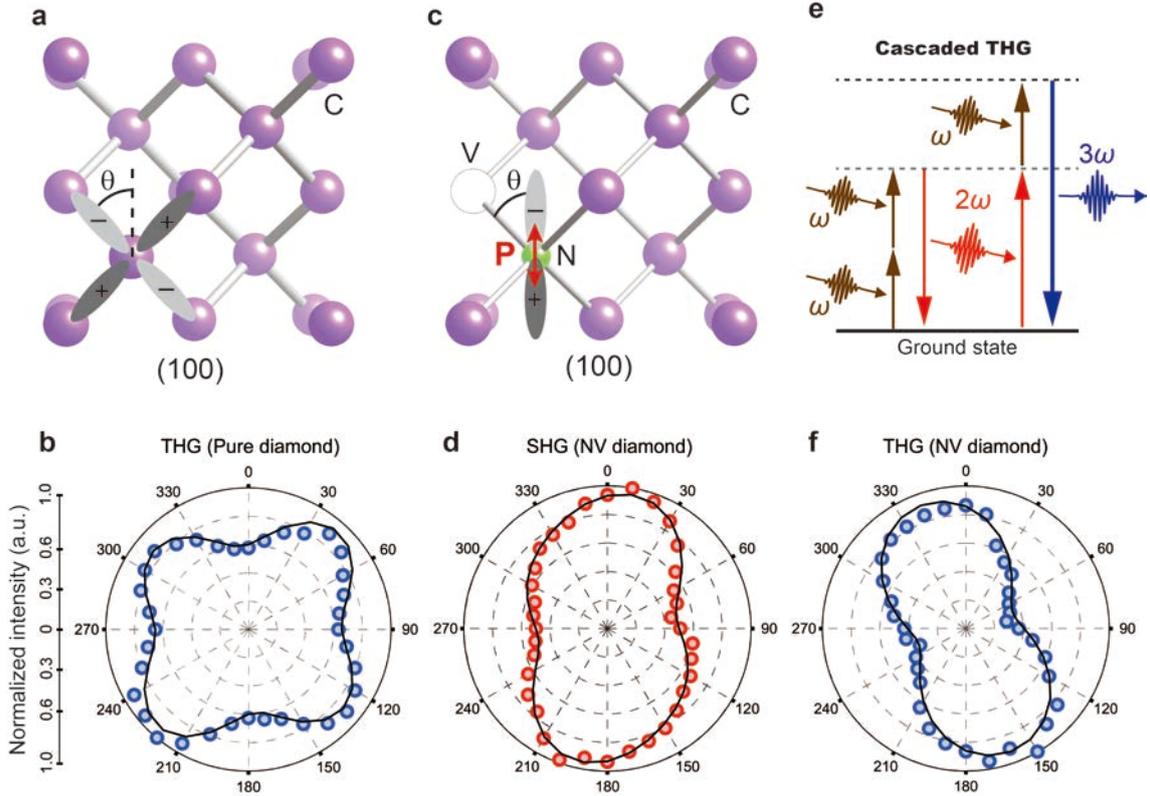

**Figure 4. Polarization dependence of nonlinear emissions from diamond crystals.** (a) The crystal lattice of pure diamond. $\theta$ is the angle of the excitation light polarization from the <001> axis. The orbitals labelled by "+" and "−" represent the quadrupole. (b) The polar plot of the normalized THG intensity with respect to the polarization angle (degree) of the excitation IR light (1350 nm, 7.6 mJ/cm$^2$) in pure diamond. (c) Same as (a), but for the NV diamond. The orbitals labelled by "+" and "−" represent dipole **P**, also indicated by the both-side arrow. (d) The polar plot of the normalized SHG intensity for the NV diamond. (e) Simplified conversion scheme of the cascaded THG process from the NV diamond. The dashed lines represent intermediate states. (f) The polar plot of the normalized cascaded THG from NV diamond crystal. The black solid curves in (b), (d), and (f) are the fit using sinusoidal functions, $A\sin 2\theta + B\sin 4\theta$, where the amplitudes A and B are the fitting parameters and $\theta$ is the angle of the light polarization with respect to the <001> axis.

In summary, we have established the second-order nonlinear optical effects of bulk diamond crystals using inversion symmetry-breaking by NV color center. In particular, the red-color SHG emission, which exhibited a power-law with exponents 2, is clearly observed from the NV diamond in addition to the blue-color cascaded THG emission, which exhibited a power-law with exponents 3, consistent with the fundamental nonlinear optics theory.[13] The traditional surface SHG from diamond was extremely weak,[32] preventing from the application of SHG. In contrast, the powerful SHG emission that we observed from the symmetry-breaking NV layer will open new avenues to the development of diamond-based second-order nonlinear photonics as well as quantum technologies. We envisage that a low-loss optoelectronic device based on the electro-optic effect—one of the second-order NLO processes[13]—can be developed, such as a superconducting qubit[33] and a diamond mechanical resonator[34], by which quantum information science would be benefited. Furthermore, precise temperature sensing using temperature-dependent phase matching for SHG will be realized, where the change in the diamond refractive index upon heating/cooling can be detected via the variation of the SHG intensity.

(Received: February 12, 2021.)

**ASSOCIATED CONTENT**

**Supporting Information:** The Supporting Information is available free of charge on the ACS Publications website at DOI:

IR excitation spectra; the tuning curve of nonlinear emissions; nonlinear conversion efficiency versus excitation wavelength; the spectral bandwidth of second-harmonic generation, third-harmonic generation, and excitation beam; the nonlinear emissions from the NV diamond for various excitation fluences.

**AUTHOR INFORMATION**

**Corresponding Authors**

*E-mail: a.abulikemu.fw@u.tsukuba.ac.jp (A. A.).

*E-mail: mhase@bk.tsukuba.ac.jp (M. H.).

## ORCID

Aizitiaili Abulikemu: 0000-0003-1457-7313

Muneaki Hase: 0000-0003-4242-2546

**Author contributions**

A. A. and M. H. planned and organized this project. Y. K. and T. A. prepared the NV diamond sample. A. A. and M. H. performed experiments and analysed the data. A. A., T. A., and M. H. discussed the results. A. A. and M. H. co-wrote the manuscript.

**Notes**

The authors declare no conflicts of interest.

**Funding**

Core Research for Evolutional Science and Technology program of the Japan Science and Technology (Grant Number: JPMJCR1875).

## ACKNOWLEDGMENT

We thank J. Afalla for useful discussions.